\documentclass{aa}
\usepackage{txfonts}
\usepackage{subfigure}
\usepackage{graphicx}
\def\lax    {\ifmmode{_<\atop^{\sim}}\else{${_<\atop^{\sim}}$}\fi}
\def\gax    {\ifmmode{_>\atop^{\sim}}\else{${_>\atop^{\sim}}$}\fi}
\def\kms    {\ifmmode{{\rm ~km~s}^{-1}}\else{~km~s$^{-1}$}\fi}

\def\arcmper  {\ifmmode \rlap.{' }\else $\rlap{.}' $\fi}

\def\arcsper  {\ifmmode \rlap.{'' }\else $\rlap{.}'' $\fi}
\def\arcsgper  {\ifmmode \rlap.^{s }\else $\rlap{.}^s $\fi}

\def\deg      {\ifmmode^\circ\else$^\circ$\fi}     

\def\hper     {\ifmmode \rlap.^{h}\else $\rlap{.}^h$\fi}

\def\m1       {$^{-1}$}

\def\mper     {\ifmmode \buildrel m\over . \else $\buildrel m\over .$\fi}

\def\>           {$>$}
\def\<           {$<$}

\def\simlt       {\lower.5ex\hbox{$\; \buildrel < \over \sim \;$}}
\def\simgt       {\lower.5ex\hbox{$\; \buildrel > \over \sim \;$}}

\begin{document}

\title{The AMIGA project. I. Optical characterization of the CIG catalog}
\author{Verdes-Montenegro, L.\inst{1}
\and
Sulentic, J.\inst{2}
\and
Lisenfeld, U.\inst{1,3}
\and
Leon, S. \inst{1}
\and
Espada, D.\inst{1}
\and
Garcia, E.  \inst{1}
\and
Sabater, J.  \inst{1}
\and
Verley, S.  \inst{1,4}
}

\offprints{lourdes@iaa.es}
\institute{Instituto de Astrof\'{\i}sica de Andaluc\'{\i}a, CSIC,
Apdo. 3004, 18080
Granada, Spain
\and
Department of Astronomy, Univ. of Alabama, Tuscaloosa, USA
\and
Dept. F\'\i sica Te\'orica y del Cosmos, Universidad de Granada, 
Spain
\and
LERMA - Observatoire de Paris, 61 av. de l'Observatoire, 75014 Paris, France}
\date{Received ; accepted }
\authorrunning{Verdes-Montenegro et al.}

\abstract{The AMIGA project (Analysis of the Interstellar Medium of
Isolated Galaxies) is compiling a multiwavelength database of isolated
galaxies that includes optical (B and H$\alpha$), infrared (FIR and
NIR) and radio (continuum plus HI and CO lines) properties. It involves a
refinement of the pioneering Catalog of Isolated Galaxies. 
This paper is the first in a series and begins
with analysis of the global properties of the nearly redshift-complete
CIG with emphasis on the Optical Luminosity Function (OLF) which we
compare with other recent estimates of the OLF for a variety of
environments. The CIG redshift distribution for n= 956 galaxies
re-enforces the evidence for a bimodal structure seen earlier in smaller
samples. The  peaks at redshift near 1500 and 6000km/s correspond
respectively to galaxies in the local supercluster and those in more
distant large-scale components (particularly Perseus-Pisces).  The two
peaks in the redshift distribution are superimposed on 50\% or more of
the sample that is distributed in a much more homogeneous way. The CIG
probably represents the most homogeneous local field example that has
ever been compiled. Our derivation of the CIG OLF is consistent with
other studies of the OLF for lower density environments.  This
comparison via the Schechter parameter formalization shows that: 1)
M$^*$ increases with galaxy surface density on the sky and 2) $\alpha$ shows a
weaker tendency to do the same. The CIG represents the largest and most
complete foundation for studies of isolated galaxies and is likely
as close as we can come to a field sample.
\footnote{Tables 1 and 2 are available in electronic form
at the CDS via anonymous ftp to cdsarc.u-strasbg.fr (130.79.128.5)
or via http://cdsweb.u-strasbg.fr/cgi-bin/qcat?J/A+A/}
}
\titlerunning{The AMIGA sample. I Optical properties. 
\thanks{
This research has made use of the NASA/IPAC Extragalactic Database
(NED) which is operated by the Jet Propulsion Laboratory, California
Institute of Technology, under contract with the National Aeronautics
and Space Administration.}} 
\maketitle

\section{Introduction}

The evolutionary history of galaxies is thought to be strongly
conditioned by the environment. Evidence has emerged for interaction-induced 
 emission enhancements (e.g. Sulentic 1976, Larson \& Tinsley
1978, Joseph \& Wright 1985, Bushouse 1987; Xu \& Sulentic 1991,
hereafter XS91) and interaction-driven secular evolutionary effects
(e.g. Moore et al. 1996; Verdes-Montenegro et al. 2001) in galaxies
that are members of binaries and dense groups. The observational
evidence is sometimes weak or unclear. Part of the difficulty lies in
the confusion between the roles of one-on-one interactions vs. more general
correlations with  average galaxy environmental density. Many of the
uncertainties, both of the amplitude of enhancements and the connection
between environment and parameters, reflect a lack of suitable control
samples to which interacting sample properties can be compared.
Ideally this would involve samples of isolated galaxies. Samples of isolated
pairs and compact groups provide the parameters to quantify effects of
2 and n body interactions. Isolated galaxy samples should provide the
baseline for interpreting the strength and properties of interaction-induced effects. Awareness of these two effects (one-on-one vs.  local
overdensity) is important in compiling isolated galaxy samples.

The most common reference or control samples found in the literature
can be described as either "field" or "normal". The former refer to
the most isolated galaxies while the latter refer to galaxies which
show none of the generally accepted signs of interaction-induced
activity. A field sample (e.g.  Kennicutt \& Kent 1983) might include
any galaxy not belonging to a cluster, so galaxies in pairs, triplets
and loose/compact groups would not necessarily be excluded. Normal
galaxy samples would be  defined in terms of specific parameters such
as HI content (Boselli et al. 2001) or a specified level of nuclear
activity. Study of  a selected quantity as a function of the
environment is then one way to quantify the level of environmentally
induced activity.

The alternative approach involves sample selection using an isolation
criterion.  In the case of isolated binaries this likely reduces the
interaction equation to the effects of one-on-one encounters. Studies
of isolated galaxies usually involve from 10s to 100-200 objects
(e.g.  Huchra \& Thuan 1977, Vettolani et al. 1986, Marquez \& Moles
1999,  Marquez et al.  2000, Colbert et al. 2001; Pisano et al. 2002,
Varela et al.  2004). The largest  samples of isolated galaxies in the
literature involve, in most cases, monochromatic observations of
subsamples from the Catalog of Isolated Galaxies (CIG: Karachentseva,
1973, also referred as K73 in SIMBAD and KIG in NED databases; see
$\S$~\ref{basic-sample}) (Adams et al. 1980, Haynes \& Giovanelli 1980,
Sulentic 1989, 
Young et al. 1986, XS91, Hernandez-Toledo et al. 1999, Perea et
al. 1997, Sauty et al. 2003).

Previous work suggests that small samples of isolated galaxies have
limited statistical value. Ideally we seek a sample large enough to
isolate a significant population of the most isolated galaxies. This
motivated us to use the CIG as the basis for a large, well-defined and
statistically significant multiwavelength database that can serve as a
comparison template for the study of galaxies in denser environments.
CIG galaxies were selected to be free of equal mass perturbers but
hierarchical pairs and groups could not be removed without reducing the
sample to negligible size. A large sample like CIG can be refined and
quantified in terms of degree of isolation. It can then be correlated
with multiwavelength interstellar medium (ISM) properties.  The result
can be a sample large enough to characterize the low density tail of
the two-point correlation function. The result  will also tell us: a)
if truly isolated galaxies exist, b) in what numbers and c) the
environmental level where the onset of interaction-induced activity can
first be detected. This study constitutes the AMIGA project (Analysis
of the Interstellar Medium of Isolated Galaxies).
 AMIGA is compiling data that will characterise 
the all phases of the ISM and it is  being released and
periodically updated at http://www.iaa.csic.es/AMIGA.html.

This paper studies the optical properties of the entire CIG as the
foundation sample for AMIGA. These properties can be compared with
future refinements to monitor changes and possible biases in the
statistical properties. The CIG is now almost redshift-complete
allowing study of both apparent and distance-dependent properties. We
first consider the distribution of CIG galaxies in 2 and 3 dimensions.
This allows us to decide which local large scale-components contribute
most heavily to the sample and how close the CIG comes to showing
homogeneity. We then analyze sample completeness via the V/V$_m$ test
followed by derivation of the optical luminosity function (OLF).  This is a
much more reliable derivation than previous estimates because of the
near completeness of redshift measures. Finally, we compare the OLF
derived for the CIG sample with those of other samples representing different
environments.

\section{The foundation of AMIGA: the original CIG catalog}
\label{basic-sample}

Statistical studies of isolated galaxies require a large, preselected
and morphologically diverse sample. The tendency for galaxies to
aggregate in multiple systems and clusters at all scales might suggest
that such a sample is difficult to find. However  Karachentseva (1973)
compiled the Catalog of Isolated Galaxies (CIG) which includes 1051
objects. All  of the CIG objects are found in the Catalog of Galaxies
and Clusters of Galaxies  (Zwicky et al. 1961-1968; CGCG) with $m_{pg}$
$<$ 15.7  and $\delta > -3 \deg$, $\sim$3\% of the CGCG).  Only one of the
compiled objects is not a galaxy, but  a globular cluster (CIG 781 =
Palomar 15, Nilson 1973), so the size of the sample considered in the
rest of this paper is n = 1050.  We used the CIG as our starting point
because it has a number of strengths:

\begin{itemize}

\item Size: The sample is large with n = 1050 galaxies. This means that
after refinement we will still be left with a statistically useful 
sample of several hundred galaxies.

\item Isolation: The CIG sample was assembled with the requirement that
no similar sized galaxies with diameter d (between 1/4 and 4 times
diameter D of the CIG galaxy) lie within 20d. Therefore for a CIG
galaxy with D = 3\arcmin, no neighbor with d = 12\arcmin\ may lie
within 240\arcmin\ and no companion with d = 0.75\arcmin\ may lie
within 15\arcmin.  It is immediately seen that this criterion is
superior to one with a fixed isolation in terms of the diameter of the
CIG galaxy in question. However it is also clear that dwarf companions
are not excluded. This is the familiar compromise between seeking
isolation and avoiding the background of distant unrelated galaxies.
There is no other way, in the absence of velocity data, to assemble a
reasonably sized catalog of reasonably isolated galaxies.  If one
assumes an average D = 25 kpc for a CIG galaxy and a typical ``field''
velocity V = 150 \kms\ then an approximately equal mass perturber would
require 3$\times$10$^9$ years to traverse a distance of 20d.  While CIG
likely contains many of the most isolated galaxies in the local
Universe it is not biased for galaxies in voids because we are usually
looking through the front side of the bubble of galaxies surrounding
the void. Thus void galaxies often fail the isolation requirement.

\item Complementarity: This CIG  is complemented by catalogs of galaxy
pairs (CPG, Catalog of Paired Galaxies; Karachentsev 1972), triplets
(Karachentseva et al. 1979) and compact groups (Hickson catalog of
Compact Groups, HCG; Hickson 1982; largely quartets). All of these
interacting comparison samples were visually compiled using  an
isolation criterion.  None of them  take into account more hierarchical
systems for the same reason that CIG could not do it. All avoid the
pitfalls associated with computer compilation from a magnitude-limited
catalog (i.e. selecting the brightest galaxy or galaxies in a
cluster).

\item Morphology: All morphological types are found in the CIG
including a significant local supercluster dwarf population.  The CIG
sample is large enough to permit discrimination on the basis of galaxy
type including approximately 100 non-dwarf early-type systems (see
e.g.  Aars et al. 2001; Marcum et al. 2004). It is also large enough to
survive isolation re-evaluation that may reveal many additional
interacting systems.

\item Depth: The CIG samples a large  enough volume of space to allow
us to sample the majority of the optical luminosity function (OLF). Galaxies
with a recession velocity less than 1000 \kms\ include the most isolated 
nearby dwarfs. Significant sampling at and beyond 10000 \kms\ allows 
us to also sample the extreme bright end of the OLF. 

\item  Completeness:  Previous work suggested that the CIG is 80-90\% 
complete to  m$_{Zw}$~15.0 (XS91). See also $\S$~\ref{complet}.

\end{itemize}

\section{Refinements of the sample}
\label{refinement}

  The CIG can be improved in several
ways that take advantage of the digitized sky surveys (POSS1 and
POSS2). Our two largest refinements include uniform reevaluation of
morphology and isolation degree. This is being done for the entire
sample  except for the nearest dwarf galaxy subpopulation  where
numerous sources of distances and morphologies now exist.

\subsection{Morphology and positions}
\label{morfo-rev}

\begin{figure}
\centering
\includegraphics[width=8cm]{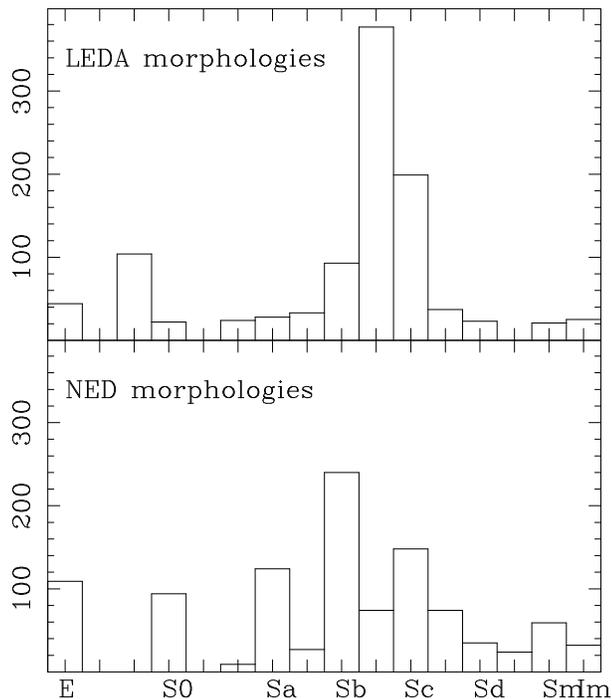}
\caption{Histogram of the morphological types of the full
CIG sample obtained from NED (bottom) and LEDA (top).}
\label{fig-morfo}
\end{figure}

The first papers discussing CIG morphology, and isolated galaxy
morphology in general, are roughly contemporary with the appearance of
the catalog in 1973. Galaxy classification data for CIG galaxies is
non-uniform and often contradictory. According to the NED and LEDA
databases 
the  CIG is composed of $\sim$ 20\% early types (E+S0), however the
distribution of the individual mophologies shows a large discrepancy,
as shown in Fig.~\ref{fig-morfo}. 
We re-evaluated CIG morphologies using the POSS2 images (Sulentic et al.
in preparation) and find it possible to obtain reliable galaxy types
for 80\% or more of the sample.  The population of luminous
isolated spirals  are the easiest to classify: near face-on spirals
could be easily recognized beyond 10000 \kms. The remaining 20\% of the
sample are being supplemented with archival data (e.g. SDSS; James et
al.  2004) or new CCD images on 1-2m class telescopes.
 POSS2 provides the higher spatial resolution necessary to distinguish
between basic subtypes.

Comparing CIG
positions in the SIMBAD database and the Updated Zwicky Catalogue (UZC;
Falco et al. 2000) we found differences of up to several tens of arcsec 
for some galaxies, large enough to make  accurate telescope pointings or cross
correlations with on-line databases impossible.  This motivated us to
systematically revise all of the CIG  positions using SExtractor on the
images of the digitized sky surveys (Leon \& Verdes-Montenegro 2003).
We found differences between old and new positions of up to
38.9\arcsec\ with a mean value of 2.4\arcsec\  for both SIMBAD and
UZC.  

\subsection{Redshift and distances}
\label{red1}

The fraction of CIG with measured redshift has almost doubled in the
past 15 years.  Our archival and bibliographic search reveals data for
almost the entire sample (956/1050 galaxies\footnote{This number is
updated in the electronic table  at http://www.iaa.csic.es/AMIGA.html
when new data become available.}). The redshift measures are given in
Table~\ref{tab:V1}.  About one half of the redshifts were found in NED
and  we compiled the rest from 37 different sources. This includes 10
new HI observations from Nancay and Green Bank (see footnote to
Table~\ref{tab:V1}).  Our search increased the CIG redshift sample by n
= 489 relative to the recent studies by  Hernandez-Toledo et al. (1999,
2001).  Redshift distances were derived for all galaxies with V $>$
1000 \kms\ and are expressed as D = V$_{3K}$/H$_0$ where V$_{3K}$ is
the velocity after the 3K correction (as given in Table~\ref{tab:V1})
and assuming H$_0$ = 75 \kms\ Mpc$^{-1}$.  3K corrected velocities are
computed in the reference frame defined by the 3K cosmological background
radiation.  They are corrected for local velocity inhomogeneities due
to the Local Group and Virgo Cluster. The velocity conversion is made
with the standard correction as defined in Courteau \& Van den Bergh
(1999).  The velocity and apex directions of the Sun relative to the
comoving frame have been derived from an analysis of the FIRAS data
(Fixsen et al.  1996) with V$_{apex} = 371 \kms$ and $(l_{apex},
b_{apex}) = (264.14\deg, 48.26\deg)$. Redshift-independent distance
estimates and references are provided for galaxies with V $<$ 1000
\kms\ (Table~\ref{tab:V2}), except for CIG 506 (V = 998  \kms), 657 (V
= 626  \kms), 711 (V = 976 \kms ), 
748 (V = 961  \kms) and 753 (V = 851  \kms) listed in
Table~\ref{tab:V1} since only redshift-derived distances could be
obtained.

\begin{table}
\caption{Recession velocities for the CIG sample (V$_r$ $>$ 1000\kms)$^{1,2}$}
\label{tab:V1}
\begin{center}
\begin{tabular}{cllc}
\hline
CIG &    V$_r$ & V$_{3K}$ & Reference$^{3}$\\
    & \kms &            &             \\
\hline
1 &    7271&        6914&             1\\
2 &    6983&        6649&             2\\
4 &    2310&        1959&             3\\
5 &    7865&        7514&             3\\
6 &    4528&        4183&             3\\
7 &   12752&       12394&             3\\
.&     .. &          ..&            .. \\
 \hline
\end{tabular}
\end{center}
\begin{list}{}{}
\item[$^{\rm 1}$] The full table is available in electronic
form at http://www.iaa.csic.es/AMIGA.html or from  CDS.
\item[$^{\rm 2}$] Data are also given for the 5 CIG galaxies
listed in $\S$ 3.2 with  V$_r$ $<$ 1000\kms.
\item[$^{\rm 3}$] (1) Thereau et al. (1998),
(2) Falco et al. (2000),
(3) Huchra et al. (1999), 
(4) De Vaucouleurs et al. (1991; RC3),
(5) SDSS early release 2001,
(6) Giovanelli \& Haynes (1993),
(7) Schneider et al. (1992), 
(8) Wegner et al. (1993),
(9) Giovanelli et al. (1997),
(10) Willick et al. (1990),
(11) Kochanek et al. (2001),
(12) Marzke et al. (1996),
(13) Strauss et al. (1992),
(14) Own data (HI spectra),
(15) Fouque et al. (1992), 
(16) Karachentsev et al. (1981),
(17) Huchra et al. (1990),
(18) NED,               
(19) Beers et al. (1995),
(20) Huchra et al. (1995),
(21) Ugryumov et al. (2001),
(22) Colless et al. (2001), 
(23) Schneider et al. (1990),
(24) Mathewson \& Ford (1996),
(25) Haynes et al. (1998),
(26) Gavazzi et al. (1999),
(27) Comte et al. (1999),
(28) Van Driel (2000),
(29) Haynes et al. (1997),
(30) Grogin et al. (1998),
(31) Kirshner et al. (1987),
(32) Young (2000),
(33) Saunders et al. (2000),
(34) Freudling et al. (1992),
(35) Haynes et al. (1999),
(36) Fisher et al. (1995),
(37) Lu et al. (1993),
(38) Pietsch et al. (1998).
\end{list}
\end{table}

\begin{table}
\caption{Distances for the CIG sample (V$_r$ $<$ 1000\kms )$^{1}$ }
\label{tab:V2}
\begin{center}
\begin{tabular}{rcc}
\hline
CIG & Distance & Reference$^{1}$\\
 & Mpc& \\
\hline
   45&   4.8 &  1\\
  105&   9.2 &  2\\
  109&  10.3 &  3\\
  112&  10.7 &  3\\
  121&   7.8 &  4\\
..&...&...\\
\hline
\end{tabular}
\end{center}
\begin{list}{}{}
\item[$^{\rm 1}$] The full table is available in electronic
form at http://www.iaa.csic.es/AMIGA.html or from  CDS.
\item[$^{\rm 2}$] (1) Karachentsev et al. (2003),
(2) Terry et al. (2002),
(3) Hamilton et al. (1996),
(4) Parodi et al. (2002),
(5) Miller et al. (2003),
(6) Tully (1998),
(7) Whiting (2003),
(8) Freedman et al. (2001),
(9) Sharina et al. (1999),
(10) Bottema et al. (2002),
(11) Garnett (2002),
(12) Karachentsev et al. (1996).
(13) Schmidt \&  Boller (1992),
(14) Karachentsev et al. (2003),
(15)  Gavazzi et al. (2000)
(16) Sofue et al. (1998),
(17) Teerikorpi et al. (1992),
(18) Solanes et al. (2002),
(19) Leonard et al. (2002),
(20) Martin (1998),
(21) Carrera et al. (2002),
(22) Papaderos et al. (1996),
(23) Swaters et al. (2002),
(24) Bottinelli et al. (1986),
(25) Bottinelli  et al. (1985),
(26) Majewski (1994),
(27) Bellazzini et al. (2002),
(28) Bottinelli et al. (1988),
(29)  Russell (2002),
(30) Bottinelli et al. (1984).

\end{list}
\end{table}

\section{Homogeneity, completeness and the optical luminosity function}
\label{rest}

\subsection{Distribution on the sky and in velocity space}

\begin{figure*}
\resizebox{14.cm}{!}{\rotatebox{270}{\includegraphics{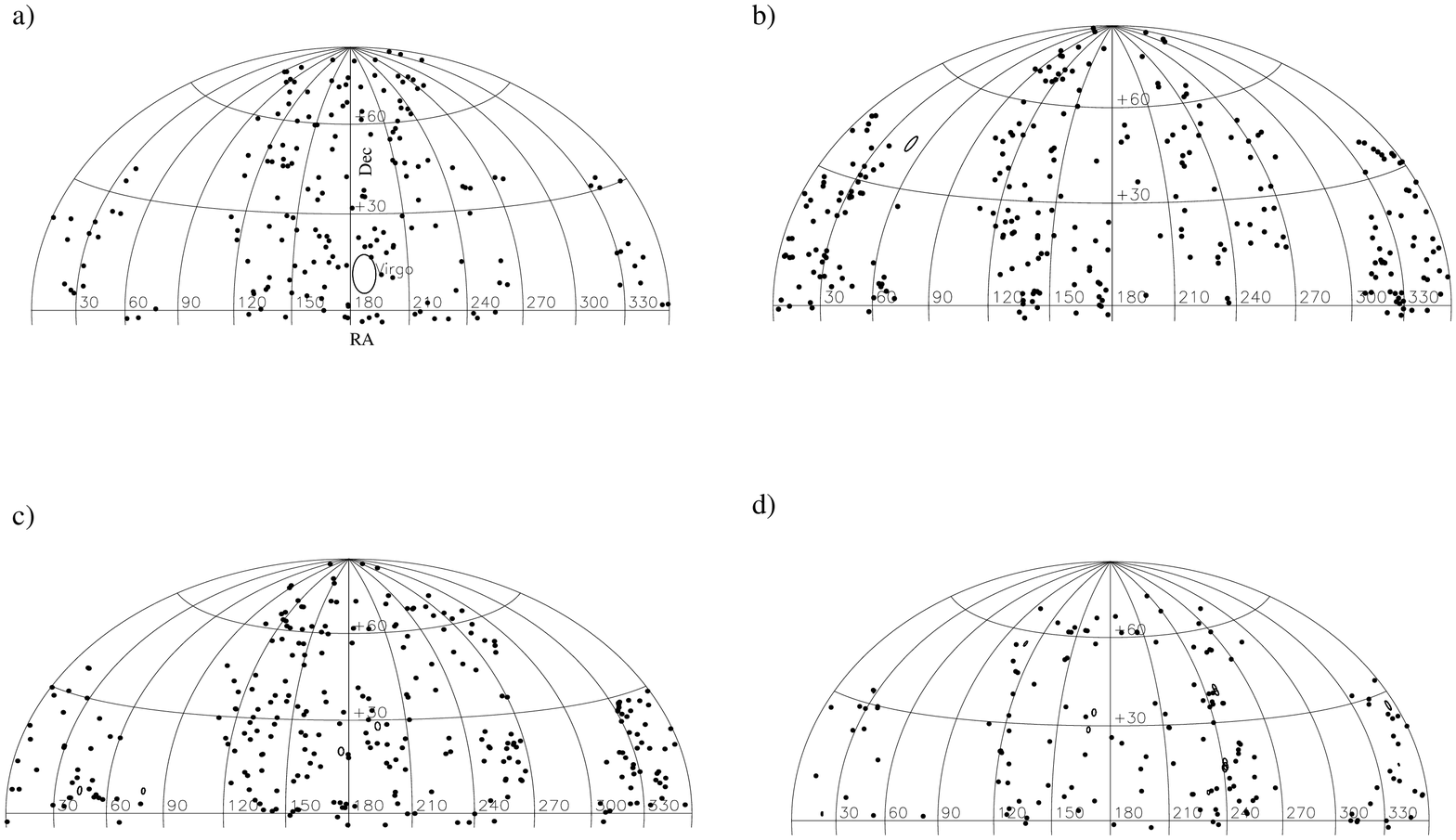}}}
\caption{Aitoff projection in right ascension and declination coordinates
showing the distribution on the sky of CIG galaxies in the 
following velocity ranges.
Galaxies in the  0-3000  \kms\ velocity interval are shown in (a), 3000 -
6000  \kms\ in (b), 6000 - 9000  \kms\ in (c) and 9000 - 12000  \kms\ in
(d). The core of the Virgo cluster is indicated by a circle with D =
12$^\circ$. Other Abell clusters in the same redshift range and with
richness classes 1 or 2 are indicated by filled circles
corresponding to their core radius.}
\label{S1}
\end{figure*}

Fig.~\ref{S1} shows the distribution of the CIG sample on the sky in
3000 \kms\ velocity intervals. This velocity segmentation makes it
easier to recognize concentrations associated with major components of
large-scale structure in the local Universe.  The core of the Virgo
cluster is indicated in the first segment with a circle of D =
12$^\circ$.  Other Abell clusters in the same redshift range and with
richness classes 1 or 2 are indicated with circles corresponding to
their core radius.  As expected  we see little correspondence between
the positions of the nearby cluster cores and CIG galaxies. Of course
some correspondence with more complex local large-scale structure
components has been found (Haynes \& Giovanelli 1983). The 2-point
correlation function for the CIG (Vettolani  et al. 1986) also shows
evidence for weak clustering.

\begin{figure*}
\centering
\includegraphics[width=10cm]{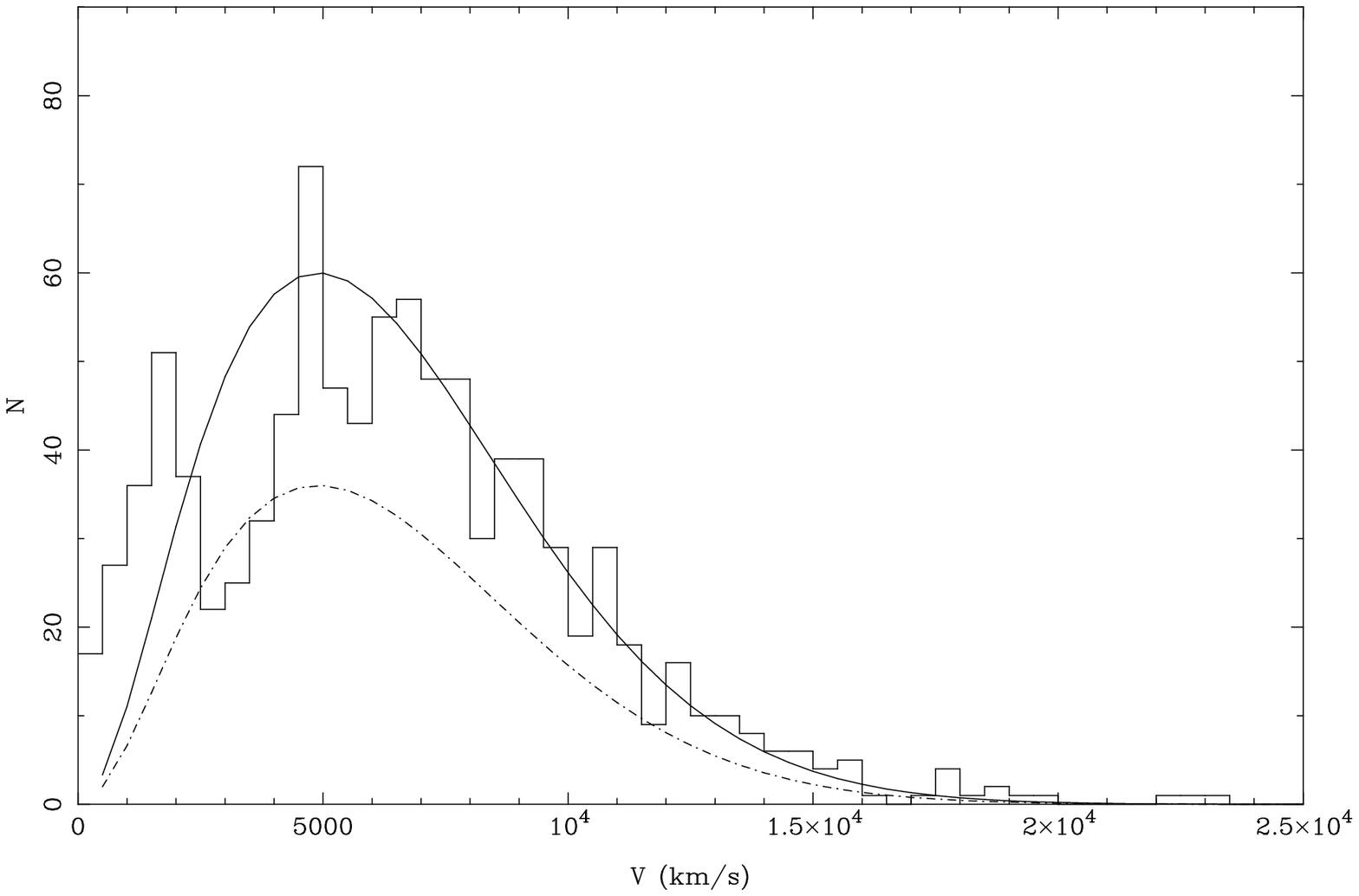}
\caption{Histogram of the optical heliocentric velocities of the 956 CIG
galaxies with redshift data. Only CIG 402 is out of the plot, with V =
40658 \kms .  The solid line corresponds to a homogeneous redshift
distribution of  the same sample size, velocity distribution and
Schechter function.  The dashed line has been obtained by scaling down
the previous distribution by a factor of 0.6.}
\label{S2}
\end{figure*}

Fig.~\ref{S2} shows the CIG redshift distribution which can be compared
with earlier studies involving smaller parts of the sample (Haynes \&
Giovanelli 1983; XS91) when far fewer redshifts were available. The
earlier studies commented on evidence for large scale structure
components including the local ($\sim$ 1500 \kms) and Pisces-Perseus
($\sim$ 5-6000 \kms) superclusters as well as the local void ($\sim$
3000 \kms) surrounding the local supercluster. It is not clear that the
latter void is actually seen. It is more appropriate to say that the
level of the curve in the 3000km/s region can be used to place an upper
limit on any quasi-homogeneous component of the CIG. It is clear that
the question of the existence of a galaxy ``field'' component is out of
date. Large-scale structure dominates the distribution of galaxies and
one can only try to isolate the lowest density regions of this
structure. Fig.\ref{S2} shows  a comparison between the CIG
redshift distribution and a corresponding homogeneous distribution of
the same sample size with the same redshift distribution and Schechter
luminosity function  (see $\S$~\ref{OLF}).  It is clear that the fit is
not satisfactory for V$<$ 6000\kms\ due to the above mentioned
structure components.  Removing these structures would provide an
estimate of the fraction of CIG galaxies that is  homogeneously
distributed, at least in 2D.  In order to estimate this number we have
assumed that the CIG is composed of both homogeneous and inhomogeneous
populations.  The latter is dominated by peaks at ~1500km/s  and
~5000km/s.  We can fit a homogeneous distribution to the population
underlying these peaks by scaling the solid curve downwards by a factor
of 0.6 (dashed curve). Thus about half of the CIG sample can be argued
to be reasonably homogeneous (see Figures 2ab). The solid curve fit to
the complete sample approaches homogeneity at $\sim$ 6500 \kms\
corresponding to a volume of a radius of about 90 Mpc). The higher velocity
part of the CIG samples a large enough volume to make sure that details
of individual large-scale structure components have little effect on
the velocity distribution. The residuals after subtraction of the
underlying homogeneous (dashed curve) component show two peaks
corresponding to the local and Pisces-Perseus Superclusters. Since the
total CIG comprises about 3\% of the CGCG  this means that ~1-2\% of
the CGCG can be argued to show homogeneity. This is about the same
population fraction as the dense isolated compact groups (Mendes de
Oliveira \& Hickson 1991, hereafter MH91; SR94) that lie at the other
end of the ``field'' clustering spectrum. This fractional similarity is
probably reasonable because both CIG (densest regions and least dense
i.e. voids) and (e.g.) HCG share a similar avoidance of the most
clustered regions via an isolation selection criterion (Sulentic 1987).
The CIG is likely as close as we can hope to come to a local
homogeneous component of the galaxy distribution.

\subsection{Optical magnitudes and completeness of the CIG sample}
\label{complet}

We compiled blue magnitudes ($m_{B}$) from the CGCG for 
all CIG galaxies and applied the following corrections. 
\begin{itemize}
 
\item Systematic errors in the CGCG catalog were reported by Kron \&
Shane (1976) (see also Giovanelli \& Haynes 1984), who showed that 
galaxies in Volume I of the CGCG had
important systematic errors relative to the rest of volumes. We applied
these corrections (A$_{v}$) to the CIG galaxies in Volume I (i.e.
galaxies with $\delta<15$ degrees and 7h$<\alpha<$18h) and with m$_{B}$
up to 15.7 mag.

\item  Galactic dust extinction (A$_{g}$) has been derived from 
IRAS/DIRBE measurements of diffuse IR emission (Schlegel 1998).

\item  Internal extinction corrections (A$_{i}$) were calculated as
a function of inclination and morphological type following RC3.
Inclinations were estimated from the ratio of major to minor axes as
given in NED. We used our revised morphologies (see
\S~\ref{morfo-rev}).

\item K corrections (Pence 1976, Giovanelli et al. 1981) were applied  
with a mean value of 0.05 mag, ranging from 0 to 0.3 mag
depending on the morphological type.
\end{itemize}

In summary, the corrected m$_{B}$ was calculated as follows:

\begin{equation}
m_{B-corr} = m_{B}+A_{v}+A_{g}+A_{i}+A_{K}
\end{equation}

We list in Table~\ref{magni} the CIG uncorrected and corrected magnitudes,
as well as the optical luminosities, derived as 

\begin{equation}
log(L_B/L_{\odot}) = 12.192 + 2 log[D(Mpc)] - 0.4 m_{B-corr}
\end{equation}

\begin{figure*}
\centering
\includegraphics[width=10cm]{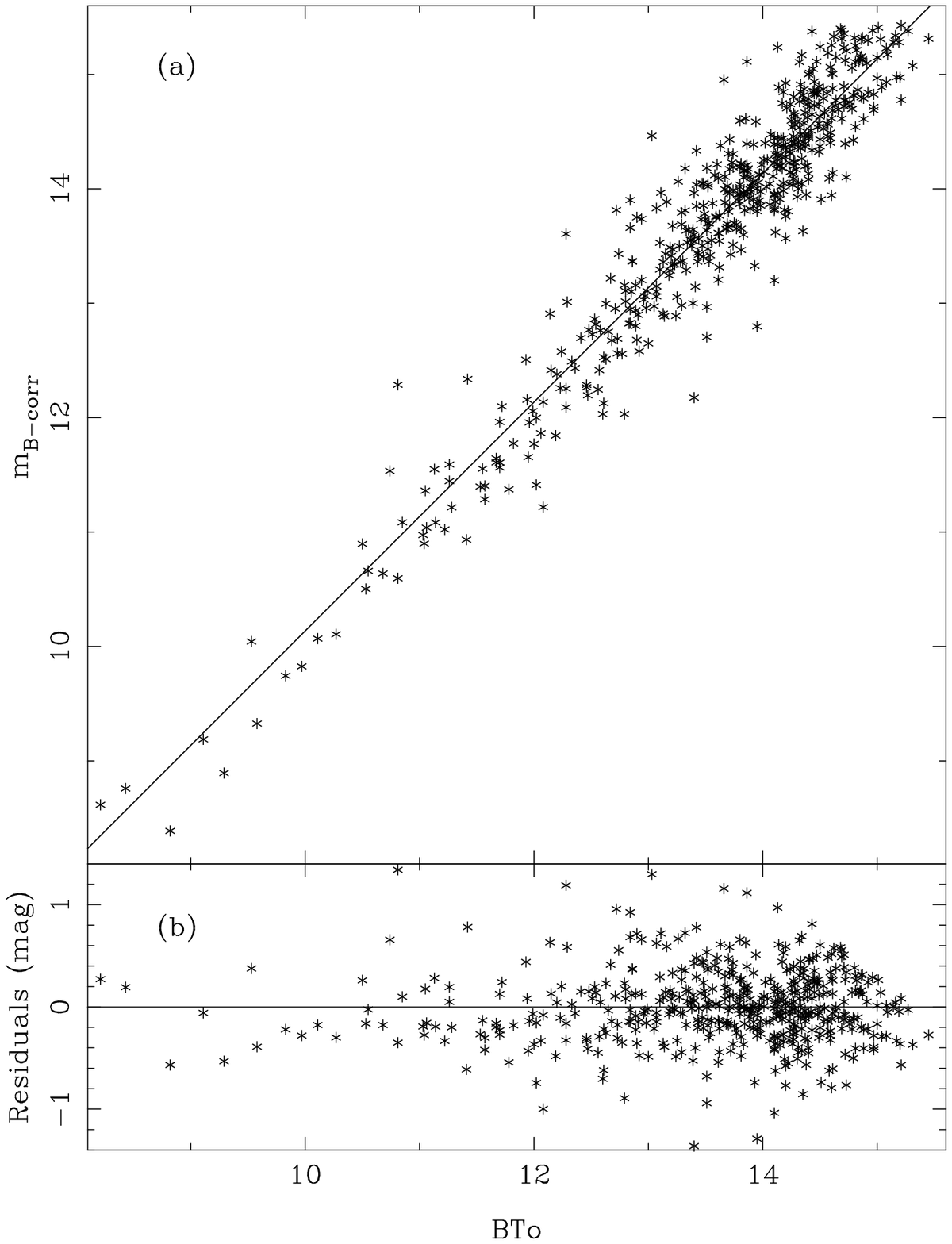}
\caption{(a) Zwicky-corrected magnitudes versus B$_T^0$ from RC3 for the 
507 CIG galaxies in common. The solid line is a fit to the plotted data and the 
parameters of the fit are given in the text ($\S$ 4.2). (b) Residuals from the 
previous fit in magnitudes versus B$_T^0$.}
\label{comp-mzw-bt0}
\end{figure*}

We compared our corrected Zwicky magnitudes m$_{B-corr}$ with 
the  B$_T^0$` values for 507 CIG galaxies found in RC3. The comparison is
shown in Fig.~\ref{comp-mzw-bt0}. A regression analysis shows that 
both quantities are linearly related as: 
\begin{equation}
m_{B-corr} = B_T^0 + 0.136 (\pm 0.001)
\label{bt0}
\end{equation}
with a correlation coefficient of 0.96. This gives confidence that 
our values are consistent with the RC3.

\begin{table}
\caption{Optical magnitudes and luminosities of the CIG sample$^{1}$.}
\label{magni}
\begin{center}
\begin{tabular}{lllr}
\hline
CIG &   m$_{B}$& m$_{B-corr}$  & L$_B$\\
    &    mag &        mag    & L$_{\odot}$ \\
\hline
1& 14.30 & 13.64& 10.67\\
2& 15.70 & 15.23& 10.00\\
3& 15.70 & 15.04& --\\
4& 12.70 & 11.55& 10.40\\
5& 15.50 & 14.52& 10.39\\
6& 14.50 & 13.69& 10.21\\
7& 15.60 & 15.30& 10.51\\
8& 15.40 & 14.18& 10.32\\
9& 15.40 & 14.54& 10.45\\
.&     .. &          ..&            .. \\
 \hline
\end{tabular}
\begin{list}{}{}
\item[$^{\rm 1}$] The full table is available in electronic
form at http://www.iaa.csic.es/AMIGA.html or from  CDS.
\end{list}
\end{center}
\end{table}

\begin{figure}
\resizebox{8.cm}{!}{\rotatebox{270}{\includegraphics{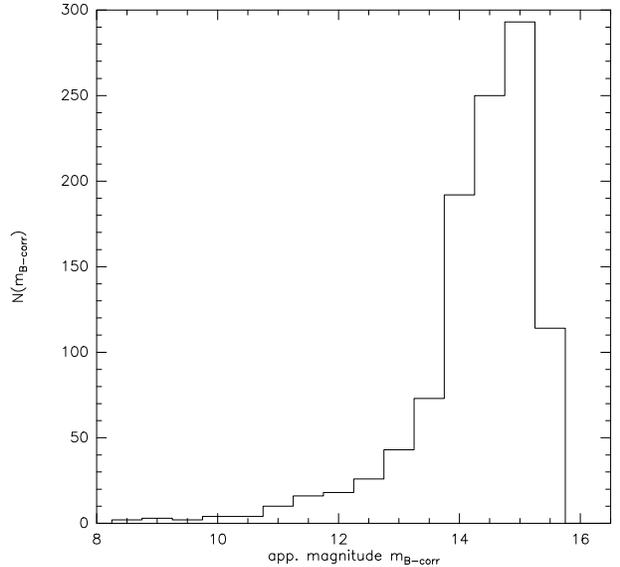}}}
\caption{Distribution of the corrected apparent magnitudes ($n = 1050$).}
\label{histo_mag_app}
\end{figure}

Figure ~\ref{histo_mag_app} shows the distribution of apparent 
corrected magnitudes for the complete sample.  We find only a few 
galaxies ($n = 19$) brighter than m$_{B-corr}$=11.0. This weak tail 
extending to m$_{app}$=8.5 represents galaxies that are in a sense 
interlopers to the CIG. They  are the  few large and bright 
galaxies in the nearby Universe that escaped deletion by the isolation 
criterion. They are almost certainly less isolated than the bulk of 
the CIG.  

\begin{figure}
\resizebox{8.cm}{!}{\rotatebox{270}{\includegraphics{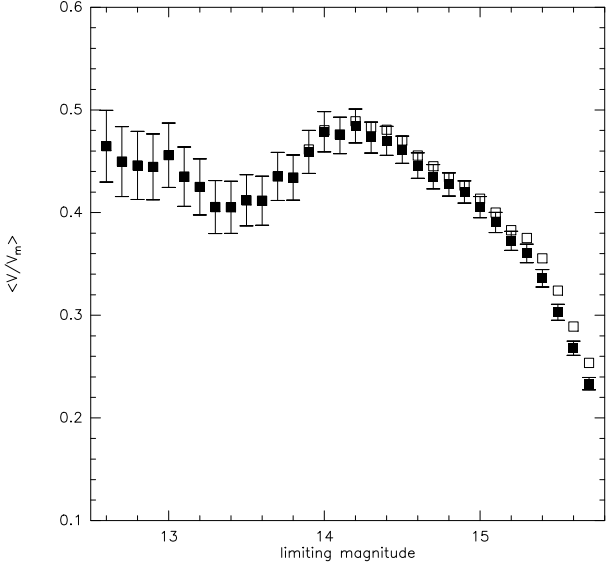}}}
\caption{The $<V/V_m>$ test for the CIG sample
excluding galaxies brighter than 11 mag. The  open squares
indicate the whole sample  ($n = 1031$), and the filled square the
subsample for which recession velocities are available ($n = 937$) and
which are used in the following to construct the OLF.  The error bars
are given for the latter subsamples and are the statistical errors in
the mean  $<V/V_m>$.  }
\label{v_vm_completeness}
\end{figure}

We used the $<V/V_m>$ test (Schmidt 1968)  to evaluate CIG sample
completeness.  We calculate for each object  the volume $V$ contained
in a sphere whose radius is the distance to the object and the maximum
volume $V_m$ contained  in a sphere whose radius is the maximum
distance at which the galaxy would still be visible given the magnitude
limit of the CIG.  We then calculate the average of the objects
brighter than the magnitude limit.  This calculation is sensitive to
the choice of the brightest magnitudes included.  We choose to neglect
galaxies brighter than 11 mag because they are obviously incomplete due
to their small number per magnitude bin (see Fig. 4). This was also
noted in the $<V/Vm>$ value when including them.  It was not necessary
to exclude galaxies fainter than 11 mag as they proved to be reasonably
complete.  Fig.~\ref{v_vm_completeness} shows the cumulative $<V/V_m>$
distribution as a function of limiting apparent magnitude. Results are
presented for the complete sample  ($n = 1031$) and the subsample for
which recession velocities are available ($n = 937$), where galaxies
brighter than 11 mag have been excluded. The difference between the two
samples is small  and only visible at fainter magnitudes where the
sample becomes slightly more complete.  Our test suggests that the CIG
is surprisingly complete (between 80-95\%) brighter than 15.0. The
sample becomes rapidly less complete at fainter magnitudes.  We
therefore adopted $m = 15.0$ ($<V/V_m> = 0.41$ for the sample with
recession velocities) as the cutoff for inclusion in the sample used to
derive the OLF. Hence the OLF has been calculated using n = 725
galaxies which corresponds to 734 galaxies with known distance and
magnitudes in the range 11 - 15 mag minus  9 galaxies with very high or
low luminosity excluded since they were scattered in bins containing a
low number of galaxies. In Table~\ref{comp_app_mag} we tabulate
$<V/V_m>$ as a function of both apparent and absolute magnitudes for
the redshift-complete sample.  The high and low luminosity bins contain
few sources because the former are rare and the latter are restricted
to the extreme low redshift part of our sample. The depression centered
at about limiting magnitude $\sim$ 13.3  reflects the gap between our
local supercluster and large-scale structure components beyond.

\begin{table*}
\caption{$<V/V_m>$ for different limiting magnitudes and different
absolute magnitude bins}
\begin{tabular}{c|cccc}
 & \multicolumn{4}{c}{Limiting apparent magnitude} \\
\hline
Absolute magnitude &  14.0       &    14.5  &         15.0  &         15.5 \\
M$_{B-corr}$ &             &          &               &              \\
\hline
  -23.0  &   0.550  (  2) &   0.276  (  2) &   0.138  (  2) &   0.069  (  2) \\
  -22.0  &   0.580  ( 15) &   0.514  ( 28) &   0.477  ( 51) &   0.291  ( 57) \\
  -21.0  &   0.536  ( 82) &   0.500  (160) &   0.426  (253) &   0.275  (296) \\
  -20.0  &   0.461  ( 79) &   0.480  (155) &   0.423  (248) &   0.315  (319) \\
  -19.0  &   0.403  ( 46) &   0.394  ( 74) &   0.344  (102) &   0.329  (146) \\
  -18.0  &   0.458  ( 25) &   0.382  ( 38) &   0.255  ( 44) &   0.232  ( 54) \\
  -17.0  &   0.540  (  7) &   0.379  (  9) &   0.465  ( 18) &   0.354  ( 25) \\
  -16.0  &   0.385  (  5) &   0.389  (  8) &   0.374  ( 11) &   0.456  ( 22) \\
\hline
 All   &   0.479  (264) &   0.461  (478) &   0.405  (734) &   0.303  (931) \\
\hline
\label{comp_app_mag}
\end{tabular}
\begin{list}{}{}
\item[~]The numbers in parenthesis give the number of galaxies in each bin ranging from
$M_{B-corr}-0.5mag$ to $M_{B-corr}+0.5mag$. All galaxies with velocity 
information and
with apparent magnitudes between 11 and and the value indicated
in the header of each column are included.
The sum of the individual bin is somewhat smaller than the total number because
the magnitudes of some galaxies fall outside the range considered.
\end{list}
\end{table*}

\subsection{CIG optical luminosity function}
\label{OLF}

The surprising level of completeness found for the CIG highlights its
strength as mentioned earlier. The completeness correction is done by
calculating  $<V/V_m>$ in intervals of 0.1 magnitude and adding the
necessary  number of  galaxies in order to obtain for $<V/V_m>$ a value
of $\sim$ 0.5, characteristic of a complete homogeneous sample (see
e.g.  Huchra \& Sargent 1973). The galaxies added in each bin are taken
into account when calculating  $<V/V_m>$  for the next fainter bin.
The final result  depends somewhat on the bin size because a change in
this parameter affects the assumed magnitude distribution of the added
galaxies.  Here we  choose to make the smallest bin size commensurate
with the precision of the data which corresponds roughly to the
expected error in the adopted apparent magnitudes.
Table~\ref{missing_galaxies} gives the results. We see that we have
added 716 galaxies in order to make  the sample complete to
m$_{app}$=15.0. This yields a correction factor $\xi$= of 2.0. We can
estimate the uncertainty of this value due to the adopted bin size. If
we had chosen a bin size of 0.2 or 0.05 mag  the number of
added galaxies would have changed to 532 and 851 respectively.  These
correction factors would have been 1.7 and 2.2 respectively.  We
estimate an uncertainty of about 15\%.

\begin{table}
\caption{Correction for incompleteness}
\begin{tabular}{ccccc}
Limiting Mag. &  $<V/V_m>$ & No. of gal. & Added gal. & Total add. gal. \\
12.100  &   0.546  &   34  &      0  &      0 \\
12.200  &   0.536  &   39  &      0  &      0 \\
12.300  &   0.532  &   45  &      0  &      0 \\
12.400  &   0.484  &   47  &      2  &      2 \\
12.500  &   0.460  &   51  &      3  &      5 \\
12.600  &   0.465  &   58  &      1  &      6 \\
12.700  &   0.450  &   63  &      3  &      9 \\
12.800  &   0.446  &   70  &      4  &     13 \\
12.900  &   0.444  &   78  &      3  &     16 \\
13.000  &   0.456  &   89  &      3  &     19 \\
13.100  &   0.435  &   96  &      8  &     27 \\
13.200  &   0.425  &  105  &      8  &     35 \\
13.300  &   0.405  &  112  &     12  &     47 \\
13.400  &   0.405  &  123  &     11  &     58 \\
13.500  &   0.412  &  137  &     11  &     69 \\
13.600  &   0.412  &  151  &     15  &     84 \\
13.700  &   0.435  &  174  &     10  &     94 \\
13.800  &   0.434  &  193  &     18  &    112 \\
13.900  &   0.459  &  226  &     11  &    123 \\
14.000  &   0.479  &  264  &     11  &    134 \\
14.100  &   0.475  &  298  &     22  &    156 \\
14.200  &   0.484  &  345  &     18  &    174 \\
14.300  &   0.473  &  384  &     34  &    208 \\
14.400  &   0.470  &  431  &     35  &    243 \\
14.500  &   0.461  &  478  &     46  &    289 \\
14.600  &   0.446  &  521  &     61  &    350 \\
14.700  &   0.435  &  570  &     70  &    420 \\
14.800  &   0.427  &  625  &     80  &    500 \\
14.900  &   0.420  &  683  &     95  &    595 \\
15.000  &   0.405  &  734  &    121  &    716 \\
15.100  &   0.390  &  785  &    143  &    859 \\
15.200  &   0.373  &  831  &    173  &   1032 \\
15.300  &   0.360  &  883  &    195  &   1227 \\
15.400  &   0.336  &  916  &    244  &   1471 \\
15.500  &   0.303  &  931  &    295  &   1766 \\
15.600  &   0.268  &  937  &    343  &   2109 \\
15.700  &   0.233  &  937  &    393  &   2502 \\
15.800  &   0.203  &  937  &    444  &   2946 \\
15.900  &   0.177  &  937  &    501  &   3447 \\
16.000  &   0.154  &  937  &    566  &   4013 \\
\hline
\end{tabular}
\label{missing_galaxies}
\end{table}

\begin{table}
\caption{Optical luminosity function}
\begin{tabular}{ccc}
\hline
M$_{\rm Zw}$ &  $\Phi$(Mpc$^{-3}$ mag$^{-1}$) & n \\   
\hline
-16.25  &  3.69E-03  $\pm$  1.26E-03  &    9 \\
-16.75  &  2.04E-03  $\pm$  6.22E-04  &   11 \\
-17.25  &  8.20E-04  $\pm$  3.16E-04  &    7 \\
-17.75  &  9.06E-04  $\pm$  2.33E-04  &   16 \\
-18.25  &  7.50E-04  $\pm$  1.43E-04  &   28 \\
-18.75  &  4.90E-04  $\pm$  8.46E-05  &   35 \\
-19.25  &  4.64E-04  $\pm$  5.76E-05  &   67 \\
-19.75  &  2.88E-04  $\pm$  3.19E-05  &   85 \\
-20.25  &  2.88E-04  $\pm$  2.30E-05  &  163 \\
-20.75  &  1.40E-04  $\pm$  1.15E-05  &  154 \\
-21.25  &  4.65E-05  $\pm$  4.76E-06  &   99 \\
-21.75  &  1.05E-05  $\pm$  1.61E-06  &   44 \\
-22.25  &  8.84E-07  $\pm$  3.39E-07  &    7 \\
\hline
\end{tabular}
\label{lum_function_tab}
\end{table}

The differential OLF $\Phi(M)$ estimating the number
of galaxies per unit volume and per unit absolute magnitude level,
is estimated from (Felten 1976)
\begin{equation}
\Phi(M)= \frac{4\pi}{\Omega} \frac{\xi}{\triangle M} \sum_i
\frac{1}{V_m(M_i)}
\end{equation}
where $\Omega$ is the sky coverage of the sample (4.38 sr for the
CIG sample from XS91)
and ${V_m(M_i)}$ is the maximum volume within which a source of
absolute magnitude $M_i$ could have been detected in a survey
down to $m_{\rm lim}$, the limiting magnitude of the sample (here: 15 mag).
The summation is over the luminosity interval $M_i+0.5\triangle M\ge M\ge
M_i-0.5\triangle M$. We have chosen $\triangle M = 0.5$.
$\xi$ is the correction factor for incompleteness derived above.
We give the so obtained OLF in Table~\ref{lum_function_tab}.
The variance of $\Phi(M)$ is estimated from
\begin{equation}
\sigma^2 = \left(\frac{4\pi}{\Omega} \frac{\xi}{\triangle M}\right)^2\sum_i
\frac{1}{V_m^2(M_i)}
\end{equation}

\begin{figure}
\resizebox{8.cm}{!}{\includegraphics{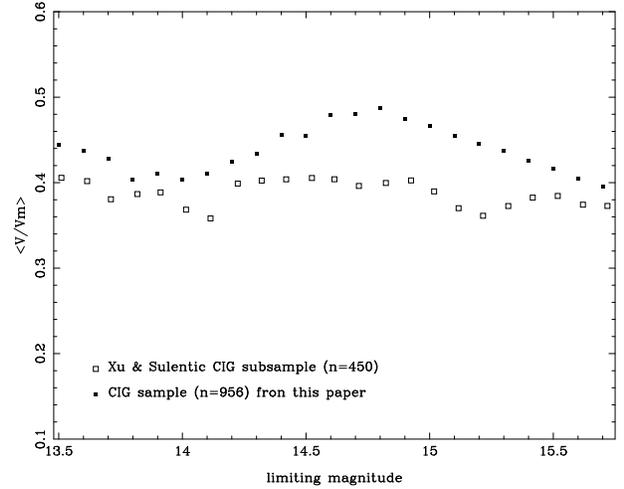}}
\caption{The $<V/V_m>$ test for our CIG sample with available
distances (n=956)
and for XS91 (n=450) sample are shown
respectively with filled and open squares. Our sample 
has been shifted by 0.5 mag toward fainter values in order to match
the uncorrected
optical magnitudes used by XS91.}
\label{v_vm_compar}
\end{figure}

Two previous estimates of the ``field'' OLF have been made using the
CIG (both given in  XS91 but partially revised in SR94). The first
involved n=295 galaxies from the Arecibo sample (hereafter AIG; Haynes
\& Giovanelli 1984) and the second involving virtually all late-type
CIG galaxies with available redshift up to 1990 (n=450). Both used
uncorrected Zwicky magnitudes. The SR94 revision the XS91 OLF
transformed the photographic magnitudes to the de Vaucouleurs
B$_T$-system with corrections for internal and external extinction. Our
derivation has two advantages: a) a 2-3$\times$ larger and reasonably
complete sample, as well as b) the ability to make more reliable
magnitude corrections using revised morphologies.  The average
magnitude correction falls in the range 0.5-0.6 magnitude and shows up
in a comparison of $<V/V_m>$ tests between our sample and previous
analyses in XS91. In Fig.~\ref{v_vm_compar}  we compare $<V/V_m>$ for
our sample shifted by 0.5 mag toward fainter values with  $<V/V_m>$
from XS91.  We find the largest difference in the range 14.5-15.5:
this is the magnitude range where most of the 400+ new redshifts
obtained in the past 10+ years are concentrated. $<V/V_m>$  appeared
too flat above 15.0 in previous evaluations.  Our derivation shows that
many of these galaxies were actually brighter than 15.0. Our $<V/V_m>$
derivation (Fig.~\ref{v_vm_completeness}) shows a more natural decline
towards fainter magnitudes.

\begin{figure}
\resizebox{8.cm}{!}{\rotatebox{0}{\includegraphics{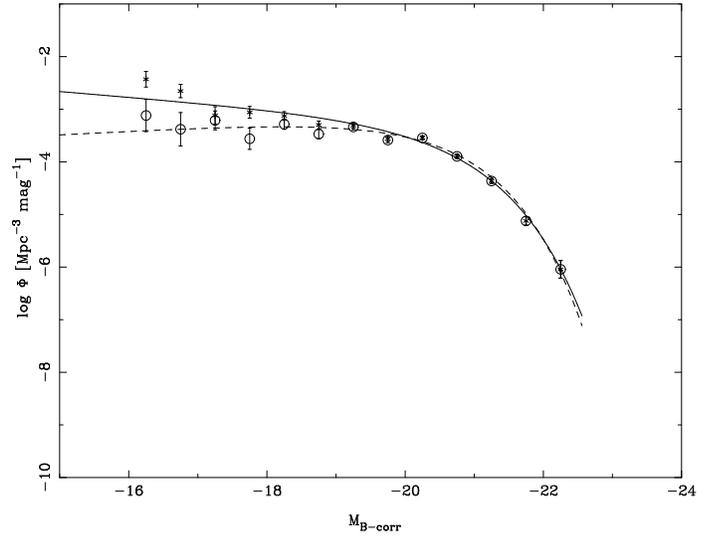}}}
\caption{Optical luminosity function for CIG galaxies  for which 
velocity information exists
and  with apparent magnitudes
between 11 and 15 ($n = 725$). The Schechter fit to this sample is 
plotted as a solid line. The 
dashed line corresponds to a fit to the same sample when 
galaxies with $V_r$ $<$ 1500 \kms\ are removed.}
\label{lum_function_fig}
\end{figure}

\begin{figure}
\resizebox{8.cm}{!}{\rotatebox{0}{\includegraphics{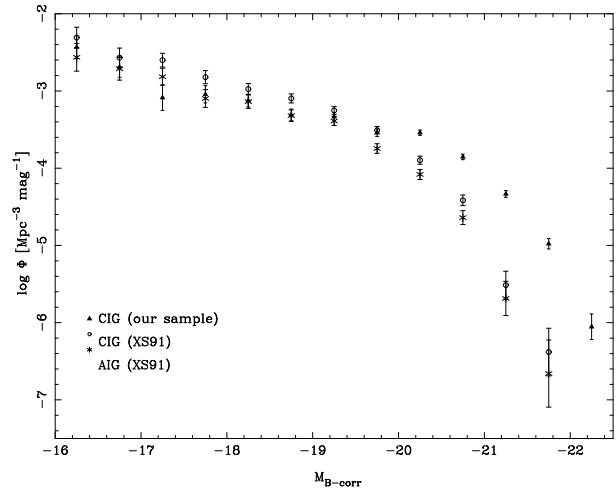}}}
\caption{All three OLF estimations for the CIG involving n=725 galaxies 
(our sample, filled triangles), 450 galaxies 
(CIG sample of XS91, open circles) and 294 galaxies 
(AIG sample of XS91, crosses). The CIG data are given in corrected
magnitudes while the data of XS91 refer to 
uncorrected magnitudes.}
\label{olf_comp}
\end{figure}

\begin{figure}
\resizebox{8.cm}{!}{\rotatebox{0}{\includegraphics{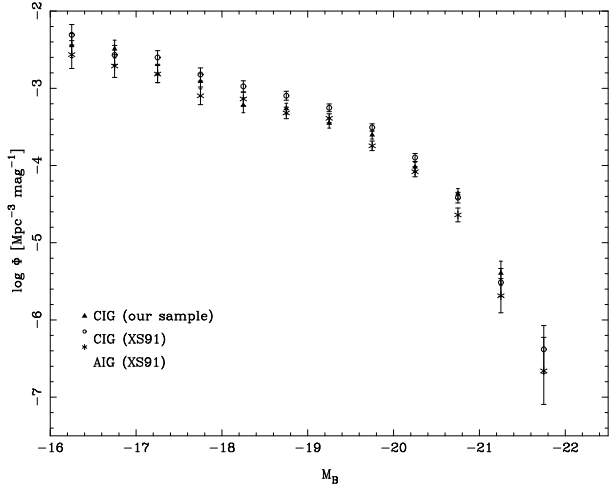}}}
\caption{The same as in Fig.~\ref{olf_comp} but calculated using uncorrected
magnitudes for the CIG galaxies, as was done for the other two samples by XS91.}
\label{olf_comp-uncorr}
\end{figure}

We fit the OLF with a Schechter function: \begin{equation} \Phi(M)=
\Phi_*10^{0.4(\alpha +1)(M^*-M)}exp(-10^{0.4(M-M^*)}) \end{equation}
using n = 725 galaxies with known distance and magnitudes in the range
11-15, once having excluded 9 galaxies with very high or low luminosity
scattered in bins containing a low number of galaxies.  The fit is
shown in Fig.~\ref{lum_function_fig} and the parameters are detailed in
Table~\ref{olf-sch-tab} where M$^*$ is given in corrected Zwicky
magnitudes (see $\S$~\ref{magni}).  The dip in our OLF at M$_B$$~$ -18
is likely related  to the dips in the redshift distribution
(Fig.~\ref{S2}) and $<V/V_m>$. All are related to the lack of
homogeneity in our sample involving the local supercluster surrounded
by a void and more distant structures.  Our next step involved
reproducing the two previous OLF determinations for the CIG indicated
above.  Figure~\ref{olf_comp} presents a comparison of all three OLF
derivations involving n=294 (AIG), 450 (CIG, XS91)  and 725 (CIG, our
sample) galaxies  respectively. We give all relevant fit parameters in
Table~\ref{olf-sch-tab} where M$^*$ for AIG and CIG (XS91) are given
in  uncorrected Zwicky magnitudes.  We transformed  M$^*$ from CIG
(SR94) to Zwicky corrected magnitudes (SR94 magnitudes were not K
corrected, a small effect of the order of 0.1 mag).  The main
difference oin our result (in agreement with SR94) with respect to XS91
is that our M$^*$ is $\sim$ 0.5 mag brighter.
 This difference is primarily due to the magnitude corrections that we
applied.  Without this correction we were able to reproduce both of the
previous OLF estimations within the error bars
(Fig.~\ref{olf_comp-uncorr}).  The small difference between our results
and those of SR94 can be attributed to the absence of K-correction in
the latter.

The $\alpha$ parameter shows less difference to previous estimates.
The underfit at the faint end results from the ``interloper''
population of dwarf galaxies mentioned earlier. All of the galaxies in
the faintest two bins (n = 21) lie within a recession velocity
V$_r$=2000  \kms\ and 13 of them within V$_r$=1000 \kms.  Only a very
small local volume is sensitive to such intrinsically faint galaxies.
However many were found in this volume  because of the ineffectiveness
of the isolation criterion for galaxies within a few Mpc.  We
recalculated the OLF without local galaxies using several velocity cuts
in addition to the restriction in magnitude range to 11-15. When we
remove galaxies with V$_r$$<$ v$_{cut}$ \kms , where v$_{cut}$ ranges
from 500 to 1500 \kms, $\alpha$ changes systematically from -1.3 to
-0.8.  Since all except 3 of the n = 59 removed galaxies 
with V$_r$ $<$ 1500 \kms\
 are fainter than M$_{B-corr}$ = -20 there is no change
in the OLF for magnitudes brighter than -20.  However the fit to the
bright part  of the OLF changes because we are truncating our
reasonably complete sample, resulting in a failure of the Schechter
function to model the OLF properly. The only interesting result to
emerge from this truncation involves the decrease in the $\alpha$
parameter.  The fit obtained for v$_{cut}$ = 1500 is shown with a
dashed line in Fig.~\ref{lum_function_fig} and the fit parameters are
given in Table~\ref{olf-sch-tab}.

\small
\begin{table*}
\caption{Optical luminosity function for the CIG sample}
\begin{tabular}{lcccc}
\hline
Sample &  $\Phi$(Mpc$^{-3}$ mag$^{-1}$) & $\alpha$& M$^*$ & Mag range for the fit\\
\hline
CIG n = 725 &                    6.3($\pm$ 0.7) $\times$ 10$^{-4}$& -1.27  $\pm$ 0.06& -20.31 $\pm$ 0.07& -16.3 to -22.3\\
CIG n = 666  & 7.5($\pm$ 0.6) $\times$ 10$^{-4}$& -0.82 $\pm$ 0.09& -20.11 $\pm$ 0.07& -16.3 to -22.3\\
(V$_r$ $<$ 1500 \kms\ excluded) & & & & \\
AIG (XS91) n = 280 &             4.7 $\times$ 10$^{-4}$             & -1.4             & -19.5$^a$&  -16 to -21.8         \\
CIG (XS91) n = 450 &            10.3 $\times$ 10$^{-4}$             & -1.4             & -19.6$^a$&   -16 to -21.8     \\
CIG (SR94) n = 450 &            8.86($\pm$ 1.51) $\times$ 10$^{-4}$ & -1.42    $\pm$ 0.08    & -20.03 $\pm$ 0.10$^b$&   -15.25 to -21.8     \\
CIG (SR94) M$\le$ -18.0 &       1.11($\pm$ 0.20) $\times$ 10$^{-3}$ & -1.20   $\pm$ 0.08 & -19.89 $\pm$ 0.12$^b$&   -18 to -21.8     \\
\hline
\end{tabular}
\label{olf-sch-tab}
\begin{list}{}{}
\item[$^{\rm a}$] Uncorrected Zwicky magnitudes. The equivalence between these 
M$^*$ and our
values for the CIG is explained in $\S$\ref{OLF}. 
\item[$^{\rm b}$] K correction was not applied to these magnitudes.
\end{list}
\end{table*}

\normalsize

\subsection{Comparison of the optical luminosity function
of the CIG with other samples in the bibliography}

In this section we  compare our CIG OLF with results of other samples 
involving a range of environments  (Table~\ref{olf-sch-tab2}). We have 
also included the results from  the compilation of Binggeli et al. (1988) 
for the OLF of field galaxies. We selected several 
samples for comparison and concentrate on the shape of the OLF rather than 
the actual space density ($\Phi$) which is an extremely difficult thing to 
compare.  The absolute magnitude range covered by each sample is given in 
Table~\ref{olf-sch-tab2}.  All published values of M$^*$ have been 
reduced to H$_0$ = 75 \kms\ Mpc$^{-1}$ and transformed to Zwicky 
corrected magnitudes (m$_{B-corr}$) using
the appropriate relation:
 m$_{B-corr}$ = B$_{MGC}$-0.124,
m$_{B-corr}$ = g$^*_{SDSS}$+0.276,
m$_{B-corr}$ = b$_{2dFGRS}$-0.054,
m$_{B-corr}$ = b$_{SSRS2}$-0.284,
obtained from the relation given in Eq.~\ref{bt0} combined with the
transformations from  Liske et al. (2003).

The comparison samples include:

\begin{itemize}

\item The Nearby Optical Galaxy (NOG) sample (Marinoni et al. 1999; v
$<$ 5500\kms).  They distinguish subsamples according to various group
properties  (Garcia 1993) for a total of 4025 galaxies.  Any galaxy not
included in one of the group categories is considered ``field''. Hence
their field sample likely  contains interacting pairs.  The
morphologies were compiled by Garcia et al. (1993) from RC3. Adopted
magnitudes were in the RC3 B$_T$ system.

\item The Second Southern Sky Redshift Survey (Marzke et al. 1998)
samples a larger volume (z$<$0.05) and contains n=5404 galaxies.  This
is a magnitude-limited sample without consideration of isolation
degree.  Morphological classifications come from several sources,
ranging from detailed
 to rough designations.  The (b$_{SSRS2}$) magnitude system is
calibrated with CCD photometry and defined to match the B(0) system
used in the CfA survey.

\item  The 2dFGRS  survey samples an even larger volume (Croton et al.
2004; z $<$ 0.11) and includes n=81387 galaxies.  They cover a wide
range of environmental densities, defined as the density contrast in
spheres of radius R=8~Mpc. Morphologies were divided into late and early
types based on spectral type.  Their photometric system
b$_J$ (noted here as b$_{2dFGRS}$) is  based on the response of the
Kodak IIIa-J +GG395 emulsion/filter combination with the zero point
determined from Johnson B-band CCD photometry.  De Propris et al.
(2003) determined the OLF for galaxies in 60 clusters from the 2dFGRS.

\item The CPG and HCG are samples where close encounters are likely to
dominate over effects of local galaxian density (i.e. they involve {\it
isolated} pairs and compact groups). Both  samples were selected using
criteria similar to the ones employed in compiling the CIG. The CPG
 was used as a control sample in previous determinations of the CPG OLF
(XS91; Toledo et al. 1999, Hernandez-Toledo et al. 1999) and contains
528 pairs.  Original Zwicky magnitudes were converted to the  B$_T^C$
system.  The HCG  OLF has been estimated by different  authors. MH91
and SR94 considered a sample composed of 68 HCGs while Zepf et al.
(1997) considered 17 HCGs together with galaxies in the close
environment of the groups. MH91 and SR94  find: 1)  a deficit of low
luminosity galaxies (depressed $\alpha$ disputed by Zepf et al.), and
2)  an excess of bright ellipticals and 3) near CIG-like behavior for
the spiral population. The latter result is in strong contrast to the
CPG OLF (XS91) where a significantly brighter  M$^*$ was found and
interpreted as the signature of interaction-induced star formation.

\end{itemize}

\begin{table*}
\caption{Optical luminosity function for the  samples from the bibliography}
\begin{tabular}{llcccc}
\hline
Sample & Reference& $\Phi$(Mpc$^{-3}$ mag$^{-1}$) & $\alpha$& M$^*$ & mag range\\
\hline
Field galaxies & Binggeli et al. &              & -0.9 to -1.25    & -19.56 to -20.36 & $\le$ -15.5 to $\le$ -18.5      \\
2dFGRS all & Croton et al. (2004)&  8.95 ($\pm$ 0.05) $\times$ 10$^{-3}$       & -1.05  $\pm$ 0.02           &-20.33  $\pm$ 0.02 & -17.7 to -22.7\\
2dFGRS void &  ''&  1.32 ($\pm$ 0.56) $\times$ 10$^{-3}$       & -1.06  $\pm$ 0.24           &-19.52 $\pm$ 0.16 & \\
2dFGRS mean & ''&  9.62($\pm$1.0) $\times$ 10$^{-3}$       & -0.99  $\pm$ 0.04           &-20.12  $\pm$ 0.05&            \\
2dFGRS cluster & ''&  25.5($\pm$ 13.2) $\times$ 10$^{-3}$  & -1.33  $\pm$ 0.11 &-20.76 $\pm$ 0.13 &            \\
2dFGRS cluster & De Propris et al. (2003)&   & -1.28  $\pm$ 0.03 &-20.75 $\pm$ 0.07 & -15.7 to -23.2           \\
SSRS2 & Marzke et al. (1998)       & 5.4($\pm$ 2.0) $\times$ 10$^{-3}$  & -1.12 $\pm$ 0.05 &-20.33 $\pm$ 0.06& -14.9 to -22.9   \\
NOG All   & Marinoni et al. (1999) & 5.9($\pm$ 0.9)  $\times$ 10$^{-3}$ & -1.10  $\pm$  0.06 &-20.53  $\pm$ 0.08 & -15.2 to -22.5   \\
NOG Field & ''                    &     & -1.19 $\pm$  0.10 &-20.45  $\pm$ 0.12 &    \\
NOG Groups& ''                    &    & -1.02 $\pm$  0.07 &-20.63  $\pm$ 0.10 &    \\
NOG Groups (n $>$ 10)& ''         &    & -1.21 $\pm$  0.11 &-20.85  $\pm$ 0.18 &   \\
NOG Groups (n $>$ 20)& ''         &    & -1.28 $\pm$  0.18 &-20.86  $\pm$ 0.31 &     \\
CPG& SR94                         & 2.60($\pm$0.24) $\times$ 10$^{-4}$& -0.90 $\pm$ 0.09& -20.34 $\pm$ 0.06\\
CPG M$\le$18.0& SR94                         & 2.31($\pm$0.22) $\times$ 10$^{-4}$& -1.06 $\pm$ 0.07& -20.24 $\pm$ 0.07\\
HCGs& MH91 &       0.55$^{+0.2}_{-0.8}$ $\times$ 10$^{-4}$   & -0.2 $\pm$  0.9    & -20.11  $\pm$ 0.20           \\
HCGs& SR94 all&  1.82($\pm$0.33) $\times$ 10$^{-5}$& -1.13 $\pm$ 0.12& -20.09 $\pm$ 0.11\\
HCGs & SR94 M$\le$18.0 & 9.21($\pm$ 2.72) $\times$ 10$^{-6}$& -1.69 $\pm$ 0.13& -20.53 $\pm$ 0.15\\
HCGs& Zepf et al. (1997)           &    &  -0.80 $\pm$  0.15 &-19.99  $\pm$ 0.16 & -14.9 to -21.9   \\
\hline
\end{tabular}
\label{olf-sch-tab2}
\end{table*}
\normalsize

\begin{figure}
\resizebox{8.cm}{!}{\rotatebox{0}{\includegraphics{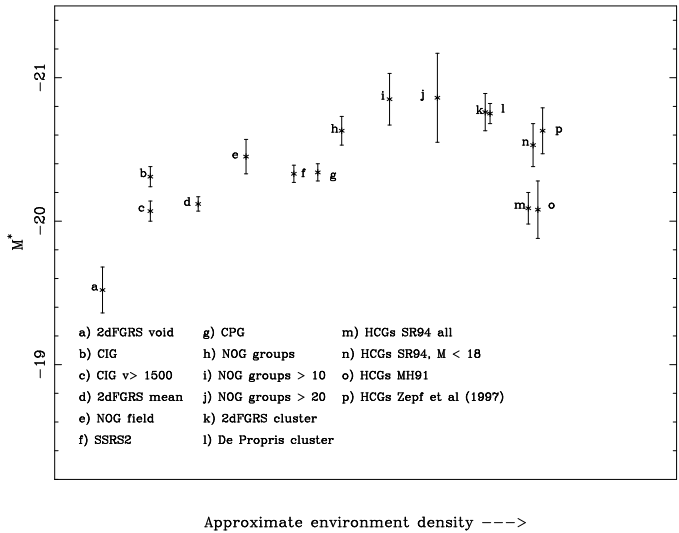}}}
\caption{The Schechter function M$^*$ parameter as a function
of the environment, ordered in an approximate way.
The label ``all'' indicates that the M$^*$ value has been obtained
for all galaxies in the sample independently of the environment.}
\label{Mstar}
\end{figure}

Fig.~\ref{Mstar} plots M$^*$ for each sample ordered roughly by
environmental density. The sequence indicates reasonably clearly the
change in OLF parameters as one proceeds from higher to lower density
samples. The former show an excess of  high luminosity galaxies as
inferred from M$^*$. Our CIG value for M$^*$  betterfits 
somewhat denser  environment than voids but with a 
lower density than some field
estimates.  This is consistent with the fact that the
isolation criterion mitigates any possible bias towards inclusion of
void members in the CIG.  Void galaxies are often not isolated in
projection but only in 3D.  Even void  samples contain interacting
pairs (e.g. Grogin \& Geller 2000) and M$^*$ will be affected depending
on their fractional representation in the sample. 

\begin{figure}
\resizebox{8.cm}{!}{\rotatebox{0}{\includegraphics{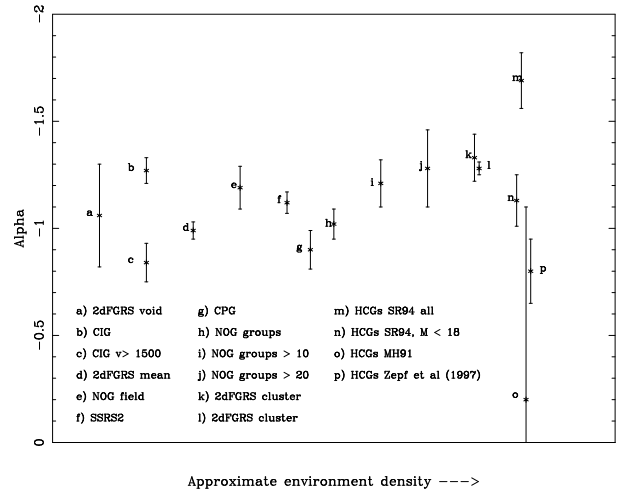}}}
\caption{The same as in Fig.~\ref{Mstar} for the $\alpha$ parameter.}
\label{alpha}
\end{figure}

We also find a possible  environmental trend (albeit with larger
scatter) for the $\alpha$ parameter (Fig.~\ref{alpha}) in the sense
that it becomes more negative for denser environments. The location of
the CIG in this plot obviously depends on the inclusion or exclusion of
the local part of the sample (V$_r$ $<$1500  \kms) which is dominated
by low luminosity dwarf galaxies (M$>$ -19). Our results are consistent
with Marinoni et al. (1999) when local dwarfs are included in the CIG
which is reasonable since they sample galaxies with  absolute
magnitudes down to M$_B$ = -15.2.  Results for the HCG are
controversial since a significant dwarf population is only found if one
increases the diameter of the groups as defined for the high luminosity
members. While low luminosity CPG pairs are found, very few compact
groups composed entirely of low luminosity members are found, for example, 
in the HCG (see SR94 Figure 1).

\section{Concluding remarks}

The CIG sample is the basis of the AMIGA project. It has many
advantages as a source of galaxies in low density environments, not the
least of which is its relatively large size. This means that it can be
refined without reducing the final sample population below a size that
would be statistically useful. We find that its 2D distribution is reasonably
homogeneous as we would expect for a distribution sampling, predominantly, the
peripheries of large-scale structure features. It is affected by the
local and Pisces-Perseus superclusters in 3D. 
The former because we are
inside it and the latter because it is rather large and diffuse.
Underlying these two bumps in the redshift distribution we again find
evidence that 50\% or more of the sample shows a quasi-homogeneous
redshift distribution, motivating us to suggest that CIG is as close as
we can hope to come towards achieving a local  ``field'' population.  A
V/V$_m$ test confirms the completeness of the CIG and a comparison of the OLF of the
CIG with that of
other samples re-enforces the credibility of the idea that CIG OLF is
representative of the lower density parts of the galaxy environment.
Care must be taken with the local supercluster contribution to the CIG
because it  samples the OLF to much lower luminosities than the rest 
of the sample.

\vfill\eject

\parindent 0pt

\begin{acknowledgements}
  LV--M, UL, DE, SL, SV and EG are
partially supported by DGI (Spain)
AYA 2002-03338
and Junta de Andaluc\'{\i}a TIC-114 (Spain).

\end{acknowledgements}

 \end{document}